%% file: main.tex
\documentclass[conference]{IEEEtran}
\IEEEoverridecommandlockouts
\usepackage{amsthm}
\newtheorem{lemma}{Lemma}

\newtheorem{theorem}{Theorem}
\usepackage{times,amsmath,epsfig}
\usepackage{epstopdf}
\usepackage{amssymb}
\usepackage{scalerel}
\usepackage{cite}
\usepackage{graphicx}
\usepackage{url}
\usepackage[bookmarks=false]{hyperref}
\usepackage{multirow}
\usepackage{multicol}
\usepackage{subfigure}
\usepackage{threeparttable}
\usepackage{booktabs}
\usepackage{ulem}
\usepackage{amsfonts}
\usepackage{textcomp}
\usepackage[framemethod=1]{mdframed}
\usepackage{enumitem}

\ifodd 0
    \newcommand\rev[1]{{\color{blue}#1}}
    \newcommand{\com}[1]{\textbf{\color{red} (COMMENT: #1)}} %comment of the text
\else
    \newcommand\rev[1]{{#1}}
    \newcommand{\com}[1]{}
\fi
\definecolor{shadecolor}{rgb}{0.878906, 0.878906, 0.878906}

\def\BibTeX{{\rm B\kern-.05em{\sc i\kern-.025em b}\kern-.08em
    T\kern-.1667em\lower.7ex\hbox{E}\kern-.125emX}}

\newcommand{\tabincell}[2]{\begin{tabular}{@{}#1@{}}#2\end{tabular}}

\title{\huge DeepOPF+: A Deep Neural Network Approach for DC Optimal Power Flow for Ensuring Feasibility
}

\author{\IEEEauthorblockN{Tianyu Zhao\IEEEauthorrefmark{1},
Xiang Pan\IEEEauthorrefmark{1}, Minghua Chen\IEEEauthorrefmark{2},
Andreas Venzke\IEEEauthorrefmark{3},
and
Steven H. Low\IEEEauthorrefmark{4}}
\IEEEauthorblockA{\IEEEauthorrefmark{1}Department of Information Engineering, 
The Chinese University of Hong Kong\\
\IEEEauthorrefmark{2}School of Data Science, City University of Hong Kong\\
\IEEEauthorrefmark{3}Department of Electrical Engineering, Technical University of Denmark (DTU)\\
\IEEEauthorrefmark{4}Department of Computing and Mathematical Sciences, California Institute of Technology
}
\vspace{-0.2in}}

\begin{document}
\begin{NoHyper}
%\settopmatter{printfolios=true}
\maketitle
\input{1abstract.tex}
\input{0notation.tex}
\input{2intro.tex}

\input{3related-work.tex}
\input{4reivew-OPF.tex}
\input{5DL.for.OPF.tex}
\input{6simulation.tex}
\input{7conclusion.tex}

%\normalem
\bibliographystyle{IEEEtran}%\vspace{-0.05in}
\bibliography{ref}

\input{8appendix.tex}
\end{NoHyper}
\end{document}

%% file: 1abstract.tex
\begin{abstract}
% MODIFICATIONS MADE BY ANDREAS VENZKE
Deep Neural Networks (DNNs) approaches for the Optimal Power Flow (OPF) problem received considerable attention recently. A key challenge of these approaches lies in ensuring the feasibility of the predicted solutions to physical system constraints. Due to the inherent approximation errors, the solutions predicted by DNNs may violate the operating constraints, e.g., the transmission line capacities, limiting their applicability in practice. To address this challenge, we develop \textsf{DeepOPF+} as a DNN approach based on the so-called ``preventive'' framework. Specifically, we calibrate the generation and transmission line limits used in the DNN training, thereby anticipating approximation errors and ensuring that the resulting predicted solutions remain feasible. We theoretically characterize the calibration magnitude necessary for ensuring universal feasibility. Our \textsf{DeepOPF+} approach improves over existing DNN-based schemes in that it ensures feasibility and achieves a consistent speed up performance in both light-load and heavy-load regimes. Detailed simulation results on a range of test instances show that the proposed \textsf{DeepOPF+} generates 100\% feasible solutions with minor optimality loss. Meanwhile, it achieves a computational speedup of \textit{two orders of magnitude} compared to state-of-the-art solvers.

\end{abstract}

%% file: 0notation.tex
\section*{Nomenclature}

%\addcontentsline{toc}{section}{Nomenclature}
{\small \begin{IEEEdescription}[\IEEEusemathlabelsep\IEEEsetlabelwidth{MMM}]
    \item[\textbf{Variable}\  \textbf{Definition}]
    \item[$\mathcal{N}$] Set of buses, $N \triangleq |\mathcal{N}|$.
    \item[$\mathcal{G}$] Set of generators. %$G \triangleq |\mathcal{G}|$.
    \item[$\mathcal{D}$] Set of loads. %$D \triangleq |\mathcal{D}| $.
    \item[$\mathcal{E}$] Set of branches.% $G \triangleq |\mathcal{E}|$.
%    \item[$\mathcal{C}$] Set of contingency cases.% $C \triangleq |\mathcal{C}| $.
    \item[$P_G$] Power generation injection vector, $[P_{G_i}, i\in \mathcal{N}]$.
    \item[$P_G^{\min}$] Minimum generator output vector, $[P^{\min}_{G_i}, i\in \mathcal{N}]$.
    \item[$P_G^{\max}$] Maximum generator output vector, $[P^{\max}_{G_i}, i\in \mathcal{N}]$.
    \item[$P_D$] Power load vector, $[P_{D_i}, i\in \mathcal{N}]$.
    \item[$\varTheta$] Voltage angle vector.
    \item[$\theta_{i}$] Voltage angle for bus $i$.
    \item[$\mathbf{B}$] Admittance matrix.
    \item[${x_{ij}}$] Line reactance from bus $i$ to $j$.
    \item[$P_{Tij}^{\max}$] Line transmission limit from bus $i$ to $j$.
    \item[$N_{{\text{hid}}}$] The number of hidden layers in the neural network.
\end{IEEEdescription}
We use $|\cdot|$ to denote the size of a set. %Note that for the buses without generators, the corresponding active generator output as well as the minimum/maximum bound of the generator output are equal to 0 i.e. $P_{G_i}=P^{\min}_{G_i}=P^{\max}_{G_i}=0, i \notin \mathcal{G}$. Similarly, $P_{D_i}=0, i \notin \mathcal{D}$.
}

%% file: 2intro.tex
\section{Introduction}\label{sec:intro}
Deep Neural Networks (DNNs) achieve superb performance in various complex engineering tasks~\cite{goodfellow2016deepma}. Their capability of approximating any continuous mapping and their scalability make DNNs a favorable choice for predicting solutions to challenging large-scale optimization problems. Motivated by this, the learning-based approaches for the OPF problem were proposed and achieve desirable performances~\cite{deepopf1,deepopf2}. %ived considerable attention~\cite{wehenkel2012automatic,li2018alphago}.
However, to ensure the feasibility of the obtained solutions to system constraints is the main challenge. For example, the existing schemes can work well in light-load regimes (i.e., the system constraints are not highly binding). However, they may generate infeasible solutions due to the inevitable approximation errors, especially in high-load regimes (i.e., the system constraints are highly binding). It may lead to an undesirable increase in computational time, as inaccurate predictions require a high-computational post-processing procedure to restore feasibility. To address this issue, we propose a preventive learning approach named \textsf{DeepOPF+}. The advantage of the proposed approach is that ensuring the feasibility of the solution without relying on a computationally expensive post-processing procedure by calibrating the system constraints (including the generation and line limits) used in training. The proposed \textsf{DeepOPF+} improves upon existing approaches~\cite{deepopf1} as it ensures feasibility in both light-load and heavy-load regimes while at the same time achieving significant computational speed-ups. 
%\rev{As \textsf{DeepOPF+} is the initial preventive learning approach to solving OPF problems, we focus on the standard DC-OPF setting in this paper.}  
The \textsf{DeepOPF+} approach can apply to more general settings, including security-constrained OPF~\cite{deepopf2} and non-convex AC-OPF problems, which we leave for future studies.

We summarize our main contributions in the following.
First, after reviewing the DC-OPF problem in Sec.~\ref{sec:OPF.review}, we propose the preventive learning framework for \textsf{DeepOPF+} in Sec.~\ref{sec: DeepOPFPlus}. Specifically, we introduce the preventive calibration of constraint limits during the training to ensure the feasibility of the predicted solutions. \rev{We remark that for each power network, we train a DNN model to approximate the corresponding load-generation mapping and predict the generations from the load inputs.} Second, as described in Sec.~\ref{ssec:ensuring-feasibility}, we provide a theoretical analysis of the relationship between the preventive constraint calibration and the approximation error of DNN. 
Finally, in Sec.~\ref{sec:simulations}, simulation results using IEEE 30, 118, 200, and 300 bus test cases show that \textsf{DeepOPF+} generates 100\% feasible solutions with minor optimality loss under suitable calibration. Meanwhile, it achieves a \textit{two orders of magnitude} computational speed-up compared to conventional approaches. 

% always generates feasible solutions with negligible optimality loss, while achieving a computational speeding-up of two orders of magnitude as compared to conventional approaches. % we carry out simulation studies using pypower \cite{tpcwTrey1} and summarize the results in
%\vspace{-0.1in}

%% file: 3related-work.tex
\section{Related Work}\label{sec: OPF.review}
Machine learning, including neural networks, has been applied to challenging power system problems for decades, %e.g.,~\cite{zhang2018real,zhang2019real,zamzam2019physics}, 
for a comprehensive review please refer to~\cite{wehenkel2012automatic}. The recent advances made in deep learning have renewed interest in applications for power systems~\cite{li2018alphago}. For brevity, we focus here on learning-based methods for solving OPF problems. The current learning-based work consists of two categories. The first category is a hybrid approach, which integrates the learning techniques into the conventional solution algorithm to solve challenging OPF %canyasse2017supervised 
problems~\cite{gutierrez2010neural,vaccaro2016knowledge,halilbavsic2018data,biagioni2019learning,baker2019learning,jamei2019meta,deka2019learning,karagiannopoulos2019data,baker2019joint,ng2018statistical,misra2018learning,zhai2010fast,roald2019implied,pineda2019data}. However, the core of these methods is still the traditional solver, which may incur high computational costs for large-scale power systems.

The second category is the end-to-end approach, which leverages machine learning models to predict solutions to OPF problems directly~\cite{deepopf1,deepopf2,guha2019machine,zamzam2019learning,fioretto2019predicting,dobbe2019towards,sanseverino2016multi}. 
The main challenge is to ensure that the predicted solutions satisfy the equality and inequality constraints. Existing works such as~\cite{deepopf1,deepopf2} introduced a post-processing procedure to handle this issue, which, however, can still be computationally expensive. \rev{To the best of our knowledge, developing the end-to-end DNN approach to solving the DC/SCDC-OPF problems is first proposed in~\cite{deepopf1,deepopf2}, where a predict-and-reconstruct framework was proposed. The work in~\cite{zamzam2019learning} applies the framework in \cite{deepopf1, deepopf2} to solve AC-OPF problems, but without considering the operating constraints on generations/line flows, which leads to a substantial fraction of infeasible solutions for the test cases. Recently, the authors in~\cite{pan2020deepopf} generalize the predict-and-reconstruct framework to the AC-OPF settings and explores a penalty approach with zero-order optimization techniques to significantly improve the feasibility of the obtained AC-OPF solutions. The zero-order optimization techniques address the challenge of not having an explicit form of the penalty terms related to the generations/line constraint violations and thus not able to apply the conventional first-order techniques like stochastic gradient decent.} Different from the existing end-to-end learning-based approaches, our proposed \textsf{DeepOPF+}  systematically calibrates constraint limits during the training stage. The proposed approach can ensure the feasibility of the obtained solutions without involving any post-processing procedure in the test stage. In our proposed \textsf{DeepOPF+}, we consider the line limits and the generation limits, add a loss term penalizing constraint violations and systematically calibrate constraint limits during training \rev{to obtain universal feasibility and consistent speedups}. We also provide a theoretical analysis of the required constraint calibration.

%% file: 4reivew-OPF.tex
\section{The DC-OPF Problem} \label{sec:OPF.review}
The DC-OPF problem~\cite{gomez2018electric} can be formulated as follows:%\vspace{-0.02in}
\begin{align}
    \min_{\mathbf{P}_{\mathbf{G}},{\varTheta }} \quad &\mathrm{\ }\sum_{i\in\mathcal{G}}{g_i\left( P_{Gi} \right)} \label{eq:DC-OPF.obj}\\%\vspace{-0.02in}
    \mathrm{s.t.} \quad& P_{Gi}^{\min}\le P_{Gi}\le P_{Gi}^{\max},\,\,\forall i\in \mathcal{G}, \label{eq:DC-OPF.generator.limit}\\%\vspace{-0.02in}
    & \mathbf{B}\cdot \varTheta =\mathbf{P}_{\mathbf{G}}-\mathbf{P}_{\mathbf{D}}, \label{eq:DC-OPF.pf}\\%\vspace{-0.02in}
    & \frac{1}{x_{ij}}\left( \theta _i-\theta _j \right) \le P_{Tij}^{\max},\,\, \forall (i,j)\in \mathcal{E}. \label{eq:DC-OPF.line.capacity}
\end{align}
\noindent %The term $N_{\text{bus}}$ is the number of buses and $N_{\text{gen}}$ is the number of generators. $P_{Gi}$ is the power output of the generator connected to the $i$th bus. $P_{Gi}^{\min}$ and $P_{Gi}^{\max}$ are the minimum and maximum generator limit, respectively. $\mathbf{B}$ is the $N_{\text{bus}} \times N_{\text{bus}}$ admittance matrix.  %They are set to 0 for the buses without generators. $
%Here $P_{Tij}^{\max}$ denotes the transmission limit for the branch connecting buses $i$ and $j$. $\mathbf{B}$ is the admittance matrix.
%which is an $N \times N$ matrix with entries
%\begin{equation}
%\mathbf{B}_c=\left[ \begin{matrix}
%B_{11,c}&		B_{12,c}&		...&		%B_{1N,c}\\
%B_{21,c}&		B_{22,c}&		...&		%B_{2N,c}\\
%...&		...&		...&		...\\
%B_{N1,c}&		B_{N2,c}&		...&		%B_{NN,c}\\
%\end{matrix} \right],
%\label{equation}
%\end{equation}
\noindent
% \[ B_{ij} =
%   \begin{cases}
%   0, & \quad \mbox{if } \ (i, j)\notin \mathcal{E}, i\neq j;\\
%     -\frac{1}{x_{ij}},      & \quad \mbox{if } \ (i, j)\in \mathcal{E};\\
%     \displaystyle\sum_{k=1,k\neq i}^{N} \frac{1}{x_{ij}}, & \quad \mbox{if} \quad i=j.
%   \end{cases}
% \]
%$\mathbf{P}_{\mathbf{G}}$ is the bus power generation vector and $\mathbf{P}_{\mathbf{D}}$ is the bus consumption vector.  $\mathbf{\theta}$ is the phase angles vector, in which $\theta_i$ represents the phase angles at the $i$-th bus. The inequalities in \eqref{eq:DC-OPF.generator.limit} are the generator limit constraints. The constraints in \eqref{eq:DC-OPF.pf} are the power flow balance equations. The constraints in \eqref{eq:DC-OPF.line.capacity} are the  transmission line capacity constraints. In the objective function, $C_i\left( P_{Gi} \right)$ refers to the cost function associated with the generator at the $i$-th bus. The cost function is derived from a heat rate curve and is commonly modeled as a quadratic function~\cite{260897}. As such, the DC-OPF problem in \eqref{eq:DC-OPF.obj}-\eqref{eq:DC-OPF.line.capacity} is a quadratic programming problem with linear constraints. % Basically, they require that total power injection into a bus should be equal to net load on the bus.

The first set of constraints in the formulation describe the generation limits. The second set of constraints are the DC power flow equations. The third set of constraints capture the line transmission capacity. In the objective, $g_i\left( P_{Gi} \right)$ is the cost function for the generator at the $i$-th bus, commonly modeled as a quadratic function~\cite{260897}: 
\begin{equation}%\vspace{-0.02in}
g_i\left(P_{Gi} \right) =\lambda _{1i}P_{Gi}^2+\lambda _{2i}P_{Gi}+\lambda _{3i}, \quad \forall i \in \mathcal{G} \label{equation5}
\end{equation}
where $\lambda_{1i}$, $\lambda_{2i}$, and $\lambda_{3i}$ are the model parameters and can be measured from data of the heat-rate curve~\cite{823997}. Noted that the DC-OPF problem is a strictly convex (quadratic) problem and thus has a unique optimal solution. {Numerical iteration solvers e.g., interior-point methods~\cite{ye1989extension} can be applied to obtain the optimal solutions. Due to the increased uncertainty from renewable generation and stochastic loads, system operators have to solve OPF closer to real-time, posing computational challenges for conventional optimization solvers, motivating the use of learning-based approaches. }  %However, the time complexity of these conventional algorithms may be substantial and depends on the transmission power system's scale~\cite{}.}
%\rev{As discussed, while the existing learning-based schemes can work well in solving DCOPF problems in light-load regimes~\cite{deepopf2}, they may generate infeasible solutions due to the inevitable approximation errors, especially in high-load regimes. This leads to an undesirable increase in computational time as infeasible predictions require a computationally complex post-processing procedure to restore feasibility. In the next section, we propose a preventive learning approach to address this issue. We note that we focus on the setting of DC-OPF in this work. The proposed preventive learning approach can be applied to more general settings, e.g., large-scale security-constrained OPF and non-convex AC-OPF problems, which guides future studies.}

%% file: 5DL.for.OPF.tex
\section{{\textsf{DeepOPF+}} for Solving DC-OPF} \label{sec: DeepOPFPlus}
\begin{figure}[!t]
	\centering
	\includegraphics[width = 0.45\textwidth]{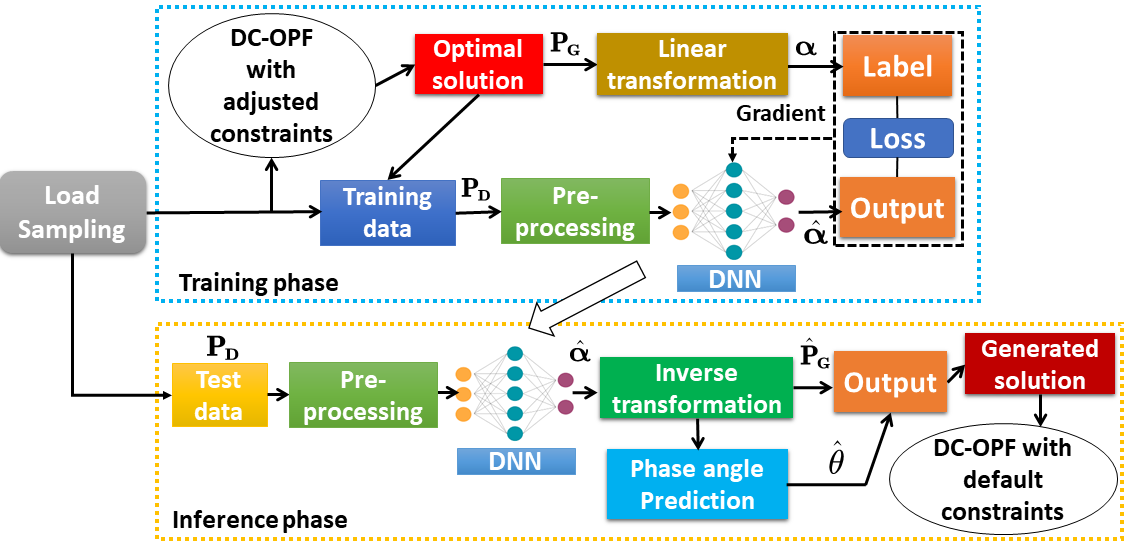}\vspace{-0.05in}
	\caption{The feasibility-ensuring learning framework of \textsf{DeepOPF+}. $\mathbf{P}_{\mathbf{D}}$ is the sampled load vector. $\mathbf{P}_{\mathbf{G}}$ ($\mathbf{\hat{P}}_{\mathbf{G}}$) is the optimal (predicted) bus power generation vector and $\mathbf{\alpha}$ ($\mathbf{\hat{\alpha}}$) is the corresponding optimal (predicted) scaling factors. $\hat{\theta}$ is the generated voltage phase angle of each bus.}
	\label{fig1}
	\vspace{-0.15in}
\end{figure}
%\vspace{-0.05in}
\subsection{Overview of \textsf{DeepOPF+}}
The methodology of the proposed \textsf{DeepOPF+} can be divided into training and inference stages. \rev{We remark that in \textsf{DeepOPF+}, we first train a DNN to learn the load-generation mapping. %predict the generations from the load inputs. 
The key novelty in \textsf{DeepOPF+} is the introduction of the feasibility-ensuring learning framework as depicted in Fig.~\ref{fig1}. We will describe the process and the analysis of \textsf{DeepOPF+} in following subsections. In the inference stage, we directly apply \textsf{DeepOPF+} to solve the DC-OPF problem with given test load inputs.}

%In the training stage, we design the learning framework to ensure the feasibility of the generated solution. Specifically, we first \rev{calibrate and reduce} the original constraint limits systematically, and collect the training data. After that, we leverage the fact that the admittance matrix (after removing the entries corresponding to the slack bus) is full rank to compute $\varTheta$ for a given $\mathbf{P}_{\mathbf{G}}$. Finally, we train a DNN model to solve the fitting problem of the scaling factor vector $\mathbf{\hat{\alpha}}$ after transformation in the second step. We train the DNN model by the data-driven stochastic gradient descent optimization algorithm. The details are in Sec.~\ref{ssec:infeasibility-prevention}.

%In the inference stage, we directly apply \textsf{DeepOPF+} to solve the DC-OPF problem with given test load inputs. We also characterize a condition that allows us to determine the calibration magnitude based on the DNN model's scale in Sec.~\ref{ssec:the-analysis}. Note that in practice, it is challenging to determine tight bounds for the calibration magnitudes. For the simulation results, we use \rev{five} illustrative constraint limit calibrations.

\subsection{Learning Framework for Ensuring Feasibility} \label{ssec:ensuring-feasibility}
\subsubsection{Constraints Calibration and Load Sampling}\label{ssec:general:idea}
\textsf{DeepOPF+} is based on a novel learning framework in which the system constraints are adjusted preventively during the training stage to ensure the feasibility of the predicted solutions during the test stage. 
The idea is that we first {calibrate} the system constraints, e.g., the transmission line and slack bus generator's output regarding capacity limits by an appropriate magnitude during the load sampling. {As discussed in Sec.~\ref{ssec:the-analysis}, the line capacity limits are reduced by a certain magnitude $\eta_{ij}$, i.e., $\frac{1}{x_{ij}}\left( \theta _i-\theta _j \right) \le P_{Tij}^{\max}-\eta_{ij},\,\, \forall\, (i,j)\in \mathcal{E}$. The slack bus generation limits should also be calibrated with $\xi$, i.e., $P^{min}_{G_s}+\xi \leq P_{G_s}\leq P^{max}_{G_s}-\xi$, where $P_{G_s}$ is the output of slack bus and $P^{min}_{G_s}$ and $P^{min}_{G_s}$ are the corresponding generation limits. Our theoretical analysis characterizes the necessary calibration magnitude for ensuring universal feasibility.} Then, we train the DNN on a dataset created with calibrated limits and evaluate its performance on a test dataset with the original limits. Thus, even with the inherent prediction error of DNN, the obtained solution can still remain feasible. 

\subsubsection{Linear transformation and mapping dimension reduction} \label{ssec:linear.transformation.and.dimension.reduction}
We first reformulate the inequality constraints on active generator power through linear scaling~\cite{deepopf1}:
\begin{equation}
P_{Gi}\left( \alpha _i \right)=\alpha _i\cdot\left( P_{Gi}^{\max}-P_{Gi}^{\min} \right)+P_{Gi}^{\min},\ \alpha _i\in \left[ 0,1 \right] ,i\in\mathcal{G},\ 
\label{equation6}
\end{equation}
%\noindent where we recall that $N_{\text{gen}}$ is the number of generators. %Thus, instead of predicting the generated power, we can predict the scaling factor $\alpha_i$ and obtain the value of $P_{Gi}$. 
Then, we leverage the fact that the admittance matrix (after removing the entries corresponding to the slack bus) is full rank to express the phase angles of all buses (except the phase angle of the slack bus) as following:
\begin{equation}
\tilde{\varTheta}=\left( \mathbf{\tilde{B}} \right) ^{-1}\left( \mathbf{\tilde{P}}_{\mathbf{G}}-\mathbf{\tilde{P}}_{\mathbf{D}} \right),
\label{equation7} 
\end{equation}
\noindent where $\mathbf{\tilde{P}}_{\mathbf{G}}$ and $\mathbf{\tilde{P}}_{\mathbf{D}}$ are the $(N-1)$-dimensional generation and load vectors for all buses except the slack bus.  {We output the $N$-dimensional phase angle vector ${\varTheta}$ by inserting a zero phase angle for the slack bus into $\tilde{\varTheta}$.} Therefore, the voltage phase angles can be inferred directly from the predicted generator set-points. As benefits, the size of the DNN model and the amount of training data and time can be reduced. 
%On the other hand, the operations in the transformation is differentiable with respect to the generated $\mathbf{\hat{P}_G}$, which makes it convenient to introduce error related to $\theta$ in the loss function.

%The transformation is to make the training easier by considering the structure of the DC-OPF problem. For one thing, we take dependency between the active power and the phase angle into account and represent the phase angle with the active power, which can reduce the number of variables to be generated. For the other, the constraints of the DC-OPF are more easier to be dealt with. After transformation, the constraints related to the generator output and the power balance can be well satisfy. Meanwhile, the penalty is introduced to represent the constraint on each transmission line. Thus, solving the DC-OPF problem changes to the fitting problem with respected to $P_{Gi}\left( \alpha _i \right)$ plus a penalty term, which means if the DNN model can approximate the mapping between the input parameter and the output, the highly accurate generated solution for the DC-OPF problem can be obtained immediately without resorting to traditional iteration based solvers. 

\subsubsection{The DNN model} \label{ssec:DNN.and.loss.function}
The DNN model is based on a multi-layer feed-forward neural network structure, which consists of a typical three-level architecture: an input layer, several hidden layers, and an output layer. 
We use Rectified Linear Unit (ReLU) as non-linear activation functions of the neurons in the hidden layers. At the last step in the output layer, the Sigmoid function is applied to project the neural network output to the interval $(0, 1)$ for the scaling factor prediction. %An example of the DNN architecture is shown in the Appendix in Fig.~\ref{fig3}. 

After constructing the DNN model, we design the corresponding loss function used in the training. The loss function consists of two parts. The first part is the sum of mean square error between each element in the generated scaling factors $\hat{\alpha}_i$ and the actual scaling factors ${\alpha}_i$ of the optimal solutions:
\begin{eqnarray}
\mathcal{L}_{PG}=\frac{1}{|\mathcal{G}|}\sum_{i\in \mathcal{G}}{\left( \hat{\alpha}_i-\alpha _i \right) ^2}.
\label{equation8}
\end{eqnarray}
%where $N_{\text{gen}}$ represents the number of generators. 
Meanwhile, we introduce a penalty term related to the inequality constraint into the loss function. We first introduce an $N_a \times N$ matrix $\mathbf{A}$ derived from the line admittance matrix~\cite{4956966}, where $N_a$ is the number of adjacent buses. Each row in $\mathbf{A}$ corresponds to an adjacent bus pair. {Given the $k$-th adjacent bus pair $(i_k,j_k)\in\mathcal{E}$, $k=1,...,N_a$, let the power flow from the $i_k$-th bus to the $j_k$-th bus. Thus, the elements, $a_{ki_k}$ and $a_{kj_k}$, the corresponding $(k, i_k)$ and $(k, j_k)$ entries of the matrix $\mathbf{A}_c$, are given as:
%\vspace{-0.15in}
\begin{equation}
    a_{ki_k}=\frac{1}{P_{Ti_kj_k}^{\max}\cdot x_{i_kj_k}}\ \mathrm{and\ }a_{kj_k}=\frac{-1}{P_{Ti_kj_k}^{\max}\cdot x_{i_kj_k}}.
\label{equation9}
\end{equation}}
{Specifically, the elements of each row in the line admittance matrix are divided by the corresponding line capacity limit, respectively.} Based on (\ref{equation7}) and (\ref{equation9}), the capacity constraints for the transmission lines in (\ref{eq:DC-OPF.line.capacity}) can be expressed as:
\begin{equation}
-1\le \left(\mathbf{A}\hat{\varTheta} \right)_k\le 1, k=1,...,N_a,
\label{equation10}
\end{equation}
\noindent where $\left(\mathbf{A}\hat{\varTheta} \right)_k$ represents the $k$-th element of $\mathbf{A}\hat{\varTheta}$. {Note
that $\hat{\varTheta}$ is the phase angle vector generated based on (\ref{equation7}), and it is computed from $\mathbf{\tilde{P}}_{\mathbf{G}}$ and $\mathbf{\tilde{P}}_{\mathbf{D}}$.}  We can then calculate $\left(\mathbf{A}\hat{\varTheta} \right)_k$. The penalty term capturing the feasibility of the generated solutions can be expressed as:
\begin{eqnarray}
\mathcal{L}_{pen}=\frac{1}{N_a}\sum_{k=1}^{N_a}{\max \left( \left( \mathbf{A}_c\hat{\varTheta}_c \right) _{k}^{2}-1,0 \right) }.
\label{equation11}
\end{eqnarray}
In summary, the loss function consists of two parts: the difference between the generated solution and the reference solution and the penalty upon solutions violating the inequality constraints. The total loss is a weighted sum of the two:
\begin{eqnarray}
\mathcal{L}_{total}=w_1\cdot \mathcal{L}_{PG}+w_2\cdot \mathcal{L}_{pen},
\label{equation12}
\end{eqnarray}
\noindent where $w_1$ and $w_2$ are positive weighting factors. The training processing can be regarded as minimizing the average value of loss function with the given training data by tuning the parameters of the DNN model, \rev{which include each layer's connection
weight matrix and bias vector.} We apply the widely-used stochastic gradient descent (SGD) method with momentum~\cite{qian1999momentum} method to update the DNN's parameters \rev{at each iteration}. \rev{We refer to~\cite{deepopf2} for details of the DNN structure and the training process of optimizing DNN's parameters.}

\subsection{Computational Complexity} \label{ssec:comp.complexity}
Recall that $N$ is the number of buses. The number of optimization variables in DCOPF, including the generations and the phase angles of all buses is $\mathcal{O}\left( N\right)$. The computational complexity of interior point methods for solving DCOPF as a convex quadratic problem is $\mathcal{O} \left( N^4\right)$, measured as the number of elementary  operations assuming that each elementary operation takes a fixed amount of time to perform~\cite{ye1989extension}. 

The computational complexity of \textsf{DeepOPF+} consists of three parts. The first is the complexity of predicting the generations using the DNN, which is $\mathcal{O} \left(N_{\scaleto{hid}{3pt}}M^2\right)$ where $M$ is the maximum number of neurons in each layer and $N_{\scaleto{hid}{3pt}}$ is the number of hidden layers in DNN. {See \cite{deepopf2} for details of the analysis.} To achieve satisfactory performance in terms of optimality loss and speed-up, we set $M$ to be $\mathcal{O} \left(N\right)$ and $N_{\scaleto{hid}{3pt}}$ to be 3. As such, the complexity for predicting the generations by our DNN is $\mathcal{O} \left(N^2\right)$.
    
The second is the complexity of computing the phase angles from the generations by directly solving (linearized) power flow equations and checking the feasibility of the results. %The process involves solving $\mathcal{O}\left(N^2 \right)$ sets of linear equations and checking the transmission line limit constraints. 
The total complexity is $\mathcal{O}\left(N^3 \right)$. 

The third is the complexity of $\ell_1$-projection, if the post-processing procedure is involved to ensure feasibility of the obtained solutions. Please refer to~\cite{deepopf2} for details of $\ell_1$-projection process. {We note that under a proper constraint limits calibration magnitude, \textsf{DeepOPF+} can always provide feasible solutions without post-processing as shown in Sec.~\ref{sec:simulations}. %However, DNN may still generate infeasible solutions if a too conservative calibration magnitude is applied, and thus, a post-processing procedure is required.
} The $\ell_1$-projection is a linear programming problem and can be solved in $\mathcal{O} \left( N^{2.5}\right)$ amount of time by using algorithms based on fast matrix multiplication~\cite{vaidya1989speeding}. 

Overall, the total computational complexity of \textsf{DeepOPF+} is $\mathcal{O}\left(N^3 \right)$.  %if the post-processing procedure is not involved, for example when the power system is operated in the light-load regime. Otherwise, it is $\mathcal{O} \left( N^{7.5}\right)$. In both cases, 
which is lower than that of the conventional interior point method, which is $\mathcal{O} \left( N^{4}\right)$. 

\subsection{Theoretical analysis} \label{ssec:the-analysis}
%In the previous part, our analysis showed that given the system load $\mathbf{P_D}$ and power generators' outputs $\mathbf{P_G}$, the power flow of the network can be uniquely determined. 
In this section, we provide a theoretical analysis on the error transfer between the prediction errors of the generator set-points obtained from DNN and the power flow {mismatch on each transmission line and the slack bus generator output.} In addition, we show that given exact bounds of the prediction errors of the generator set-points, the maximal power offsets among all lines {and the slack bus generation} can be obtained (i.e., the required magnitude by which the {line capacity} and {slack bus generation range} have to be reduced in the training stage to ensure feasibility in the test stage). {To quantify the relationship between prediction errors and power offset on lines and slack bus,} we provide the following theorem:
\begin{theorem} \label{thm:maxerror}
Let $\epsilon$ be the maximum prediction error of the DNN such that $|\hat{P}_{Gi}-P_{Gi}| \leq \epsilon$ holds for all $i=1, 2, ..., N$, where $\hat{P}_{Gi}$ is the predicted generators' output. \rev{We have 
\begin{itemize}
    \item the maximum power offsets on the $i$-th line are $k_i\cdot\epsilon$, where $k_i=%\max_{i=1, 2, ..., |\mathcal{E}|}
    \sum^{N-1}_{k=1} |M_{ik}|$, $i\in \mathcal{E}$. Thus, if the calibration for the $i$-th line capacity constraint is set to be no less than $k_i\cdot \epsilon$, $\hat{P}_G$ satisfies the line capacity constraints
    \item the maximum power offset on the slack bus generation is $(|\mathcal{G}|-1)\cdot\epsilon$, where $|\mathcal{G}|$ denotes the number of generators. Thus, if the calibration for the slack bus generation constraint is set to be no less than $(|\mathcal{G}|-1)\cdot\epsilon$, $\hat{P}_G$ satisfies the slack bus generation constraints.
\end{itemize}}  %is some constant determined by the power network topology.
\label{theorem1}
\end{theorem}

The matrix $M$ is $|\mathcal{E}|\times(N-1)$ dimensional depending on the topology of the power network {with entries of  Power Transfer Distribution Factors (PTDFs)}, and $|\mathcal{E}|$ is the number of transmission lines. Please refer to Appendix~\ref{apx:proof_of_theorem1} for a complete proof and detailed formula of $M$. {For example, the maximum $\max_{i\in\mathcal{E}}k_i=19, 52,$ $199$, and {$299$} for IEEE 30-, 118-, 200-, and the 300-cases used in our simulation, respectively.} The maximum prediction error of the DNN depends on the neural network architecture as defined in the following theorem.
%\begin{proof}

\begin{theorem} Reproduced from \cite{deepopf2}: \label{thm:worst-cast-approximation-error}
Let $\mathcal{H}$ be the class of all possible $f^*(\cdot)$ with a Lipschitz constant $\varLambda>0$. Let $\mathcal{K}$ be the class of all $f(\cdot)$, generated by a neural network with depth $N_{{\text{hid}}}$ and maximum number of neurons per layer $N_n$. Then, the maximum prediction error can be defined as:
\begin{equation}
\epsilon = \underset{f^*\in \mathcal{H}}{\max}\ \underset{f\in \mathcal{K}}{\min} \,\underset{x\in \mathcal{S}}{\max}\left| f^*\left( x \right) -f\left( x \right) \right|\geq \varLambda\cdot \frac{ d}{4\cdot(2N_n)^{N_{{\text{hid}}}}},
\label{err_bound_dnn}
\end{equation}
where $d$ is the diameter of the load input domain $\mathcal{S}$.
\end{theorem}
The theorem characterizes a lower bound on the worst-case error of using neural networks to approximate load-to-generation mappings in DC-OPF problems. \rev{Such prediction error $\epsilon$ is characterized by the DNN structure and hence can be pre-obtained before training.} {Based on Theorem~\ref{theorem1} and \ref{thm:worst-cast-approximation-error}, we derive the following lemma.}
\begin{lemma} \label{lemma1}
To guarantee the feasibility of the DNN approximating the most difficult load-to-generation mapping with a Lipschitz constant $\Lambda$, 
\begin{itemize}
    \item \rev{the capacity constraint for the $i$-th line} is required to be reduced during the training stage at least by 
    
    $k_i\cdot\epsilon = k_i \cdot \varLambda\cdot \dfrac{ d}{4\cdot(2N_n)^{N_{{\text{hid}}}}}$, $i\in\mathcal{E}$,
    \item \rev{the slack bus generation limits} are required to be calibrated during the training stage at least by
    
    $(|\mathcal{G}|-1)\cdot\epsilon=(|\mathcal{G}|-1)\cdot\varLambda\cdot \dfrac{ d}{4\cdot(2N_n)^{N_{{\text{hid}}}}}$, 
    
    %i.e., $P_{G_s}\geq P^{min}_{G_s}+(|\mathcal{G}|-1)\cdot\varLambda\cdot \dfrac{ d}{4\cdot(2N_n)^{N_{{\text{hid}}}}}$ and $P_{G_s}\leq P^{max}_{G_s}-(|\mathcal{G}|-1)\cdot\varLambda\cdot \dfrac{ d}{4\cdot(2N_n)^{N_{{\text{hid}}}}}$,
\end{itemize}
{where $P_{G_s}$ is the output of slack bus and $P^{min}_{G_s}$ and $P^{min}_{G_s}$ are the corresponding generation limits.}
\end{lemma}
The proof of Lemma~\ref{lemma1} follows directly from Theorem~\ref{thm:worst-cast-approximation-error} and Theorem~\ref{thm:maxerror}. {It indicates that the line capacity limits and the slack bus generation limits are required to be calibrated by a necessary magnitude such that even with prediction errors, DNNs could still generate feasible solutions under the worst-case.}  
Lemma~\ref{lemma1} provides the insight that larger NN sizes contribute to smaller prediction errors and thus require smaller magnitudes of reduction of system constraints to ensure universal feasibility.  {Note that in practice, it is challenging to determine tight worst-case prediction errors of DNNs.} We plan to compute the Lipschitz constants of DNNs \cite{fazlyab2019efficient} and tight bounds on the approximation error \rev{and exact constraints calibration magnitudes }in future work.  In the next section, {we show numerical results for {five} illustrative transmission line capacity limits and slack bus generation limits calibrations \rev{without prior knowledge of approximation error $\epsilon$.  }}

%% file: 6simulation.tex
\begin{table}[!t]
	\centering
	%\tiny
	\caption{Parameters for test cases.} %\rev{Recall that $N$ is the number of buses, $G$ is the number of generators, $D$ is the number of load, and $K$ is the number of branches.}}
	\renewcommand{\arraystretch}{0.9}
	\begin{threeparttable}
		\begin{tabular}{c|c|c|c|c|c|c}
			\toprule
			\hline
			Case & \tabincell{c}{$N$} & \tabincell{c}{$|\mathcal{G}|$} & \tabincell{c}{$|\mathcal{D}|$} &\tabincell{c}{$|\mathcal{K}|$} &\tabincell{c}{$N_{\scaleto{hid}{3pt}}$} &\tabincell{c}{Neurons \\per hidden layer}
			\\%&\tabincell{c}{$lr$} \\
			\hline
			\tabincell{c}{Case30} & 30 & 6 & 2 &41 & 3&32/16/8\\%&1e-3\\
			\hline
			\tabincell{c}{Case118} & 118 & 19 & 99 &186&3&128/64/32\\%&1e-3\\
			\hline
			\tabincell{c}{Case200} & 200 & 32 & 108 &245&3&128/64/32\\%&1e-3\\
			\hline
			\tabincell{c}{Case300} & 300 & 57 & 199 &411&3&256/128/64\\%&1e-3\\
			\hline
			\bottomrule
		\end{tabular}
		\begin{tablenotes}
			\footnotesize
			\item[*] The number of load buses is calculated based on the default load on each bus. A bus is considered a load bus if its default active power consumption is non-zero.
% 			\item[*] The values for these \rev{neural network} parameters are not unique. Ohjer combinations of the parameters may achieve similar performance.
		\end{tablenotes}
	\end{threeparttable}
	\label{table1}\vspace{-0.1in}
\end{table}

\begin{table*}[!ht]
	\centering
	\caption{PERFORMANCE EVALUATION OF \textsf{DeepOPF+}.}\vspace{-0.06in}
	\begin{threeparttable}
			\begin{tabular}{c|c|c|c|c|c|c|c|c|c}
			\toprule
			\hline
			\multirow{3}{*}{Case}&
			\multirow{3}{*}{\tabincell{c}{Limit \\calibration (\%)}} &
			\multirow{3}{*}{\tabincell{c}{Feasibility \\rate (\%)}} &
			\multirow{3}{*}{\tabincell{c}{Feasibility rate \\without calibration (\%)}} &
			\multicolumn{3}{c|}{\tabincell{c}{Average \\cost (\$/hr)}} &
			\multicolumn{2}{c|}{\tabincell{c}{Average running \\time (ms)}} &
			\multicolumn{1}{c}{\tabincell{c}{Average \\speedup}} \\
			\cline{5-9}
			&&&&\tabincell{c}{\textsf{DeepOPF+}}&Ref.&Loss(\%)&\textsf{DeepOPF+}&Ref.&\\ %\tabincell{c}{light-load\\regime}&\tabincell{c}{heavy-load\\regime}\\
			\hline
 			\multirow{5}{*}{Case30}&\tabincell{c}{0.5}& 94.94 &\multirow{5}{*}{94.02}& 679.0 & \multicolumn{1}{|c|}{\multirow{5}{*}{677.3}} & 0.27 &0.54 & \multicolumn{1}{|c|}{\multirow{5}{*}{44}} & $\times$85\\
  			 &\tabincell{c}{1.5}& 96.90 & & 679.0 &  & 0.27 &0.52 && $\times$86 \\
  			 &\tabincell{c}{3.5}& 100 & & 679.0 &  & 0.27 &0.50 && $\times$88 \\
  			 &\tabincell{c}{5}& 100 & & 679.0 &  & 0.27 &0.49 && $\times$88 \\
  			 &\tabincell{c}{7}& 100 & & 679.2 &  & 0.30 &0.50 && $\times$89 \\
 			\hline
 			\multirow{5}{*}{Case118}&\tabincell{c}{0.5}& 70.08 &\multirow{5}{*}{61.66}& 111617.6 & \multicolumn{1}{|c|}{\multirow{5}{*}{111219.5}} & 0.36 &1.60 & \multicolumn{1}{|c|}{\multirow{5}{*}{116}} & $\times$143\\
  			 &\tabincell{c}{1.5}& 81.12 & & 111660.8 &  & 0.40 &1.27 && $\times$152 \\
  			 &\tabincell{c}{3.5}& 97.72 & & 111752.7 &  & 0.48 &0.62 && $\times$205 \\
  			 &\tabincell{c}{5}& 100 & & 111829.6 &  & 0.55 &0.59 && $\times$200 \\
  			 &\tabincell{c}{7}& 100 & & 111925.6 &  & 0.63 &0.56 && $\times$208 \\
 			\hline
 			\multirow{5}{*}{Case200}&\tabincell{c}{0.5}& 68.58 &\multirow{5}{*}{63.36}& 39118.0 & \multicolumn{1}{|c|}{\multirow{5}{*}{38754.7}} & 0.95 &2.09 & \multicolumn{1}{|c|}{\multirow{5}{*}{104}} & $\times$114\\
  			 &\tabincell{c}{1.5}& 86.52 & & 39257.3 &  & 1.31 &1.18 && $\times$158 \\
  			 &\tabincell{c}{3.5}& 91.2 & & 39319.0 &  & 1.47 &0.96 && $\times$167 \\
  			 &\tabincell{c}{5}& 94.62 & & 39388.6 &  & 1.65 &0.81 && $\times$175 \\
  			 &\tabincell{c}{7}& 100 & & 39501.3 &  & 1.94 &0.59 && $\times$178 \\
 			\hline
 			
 			\multirow{5}{*}{Case300}&\tabincell{c}{0.5}& 78.14 &\multirow{5}{*}{75.94}& 853607.1 & \multicolumn{1}{|c|}{\multirow{5}{*}{852611.6}} & 0.11 &3.35& \multicolumn{1}{|c|}{\multirow{5}{*}{82}} & $\times$95\\
  			 &\tabincell{c}{1.5}& 86.96 & & 583581.6 &  & 0.11 &2.23 && $\times$111 \\
  			 &\tabincell{c}{3.5}& 97.92 & & 853831.8 &  & 0.14 &0.96 && $\times$118 \\
  			 &\tabincell{c}{5}& 100 & & 854187.1 &  & 0.19 &0.66 && $\times$125 \\
  			 &\tabincell{c}{7}& 100 & & 854998.5 &  & 0.28 &0.66 && $\times$126 \\
 			\hline
			\bottomrule
		\end{tabular}
		\begin{tablenotes}
			\footnotesize
 			\item[*] {If the DNN generates infeasible solutions, we apply an efficient $\ell_1$-projection post-processing procedure to ensure the feasibility of the final solution~\cite{deepopf2} by the Gurobi solver. The average running time includes the post-processing time if DNN obtains infeasible solutions.}
 			\item[*] {Speedup is calculated as the average of the running-time ratios of Pypower to \textsf{DeepOPF+} for all the test instances. We note that the speedup is the average of ratios, and it is different from the ratio of the average running times between Pypower and \textsf{DeepOPF+}.}
            \item[*] {Note that Case118 takes longer computational time to obtain the optimal solution with the conventional solver compared to Case300.  This is due to the observation that Case118 requires more iteration steps to converge (on average 25 times) than Case300 (on average 11 times), while the average running time per iteration of Case118 (4.7 ms) is less than that of Case300 (7.5 ms).}
		\end{tablenotes}
	\end{threeparttable}
	\label{table2}\vspace{-0.02in}
\end{table*}

\section{Numerical Experiments}\label{sec:simulations}
\subsection{Experiment setup}
\subsubsection{Simulation environment} The experiments are conducted in CentOS 7.6 on the quad-core (i7-3770@3.40G Hz) CPU workstation with 16GB RAM.
%Ubuntu 16.04 on the six-core (i5-8500@3.00G Hz) CPU workstation and 8GB RAM.
\subsubsection{Test case}
The proposed approach is evaluated for four representative test cases: IEEE 30-bus, 118-bus test cases~\cite{tpcwTrey4}, a 200-bus power system~\cite{birchfield2017grid}, %provided by MATPOWER ~\cite{zimmerman2011matpower}
 and IEEE 300-bus test case.\footnote{As IEEE 118-bus and 300-bus test cases provided by MATPOWER~\cite{zimmerman2011matpower} do not specify the line capacities, we use IEEE 118-bus test case provided by Power Grid Lib~\cite{babaeinejadsarookolaee2019power} and use the line capacity setting for IEEE 300-bus test case with same branch from Power Grid Lib~\cite{babaeinejadsarookolaee2019power}
(version 19.05).}% while all other parameters are taken from the MATPOWER case.}

%The topology for IEEE 30-bus system is shown in Appendix in Fig.~2 as an illustrative example. %: the IEEE 30-bus power system, IEEE 57-bus power system, the IEEE 118-bus test system and the IEEE 300-bus system, respectively, which includes  The related parameters for the test cases are shown in Table \ref{table2}. The illustrations of the topology for IEEE 30-bus system and 57-bus system are shown in Appendix as examples. The illustrations for the IEEE 118-bus and IEEE 300-bus cases can be found in ~\cite{4077121} and ~\cite{tpcwTrey6}. 

\subsubsection{Training data}
For the training stage, the load data is sampled within $[100\%, 130\%]$ of the default load on each load uniformly at random, which covers both light-load and heavy-load regimes. On the heavy-load regimes, some transmission lines and the slack bus generation will reach their upper operation limits under the given load. Based on preliminary experiments, we test \textsf{DeepOPF+} by calibrating the transmission lines' limits and slack bus generation limits by $c=$ 0.5\%, 1.5\%, 3.5\%, 5\%, and 7\%, respectively. %(i.e., $\frac{1}{x_{ij}}\left( \theta _i-\theta _j \right) \le P_{Tij}^{\max}\cdot (1-c),\,\, \forall (i,j)\in \mathcal{E}$ and $P^{min}_{G_s}+c\cdot(P^{max}_{G_s}-P^{min}_{G_s}) \leq P_{G_s}\leq P^{max}_{G_s}-c\cdot(P^{max}_{G_s}-P^{min}_{G_s})$, where $c$ is chosen as above).} 
Note that we will investigate the systematic calibration of the limits using Lemma~\ref{lemma1} in future work. The solution for the DC-OPF problem provided by Pypower~\cite{tpcwTrey1} is regarded as ground-truth. For each test case, the amount of training data and test data are 25'000 and 5'000.

\subsubsection{The implementation of the DNN model}
We design the DNN model based on the Pytorch platform. For the training process, the number of epochs is 200, and the batch size is 64. Based on the range of each loss obtained from preliminary experiments, the value of weighting factors $w_1$ and $w_2$ are each set to 1. \rev{For each power network, we train a DNN model to approximate the corresponding load-generation mapping. We remark that the DNN inputs the load profile and outputs the generation prediction. According to the size of the power network, we design DNN models with a different number of neural network layers. These parameters are given in Table~\ref{table1}.}
%The detailed architecture of the DNN model for the IEEE 30-bus test case is shown in Appendix in Fig~3. 

% \begin{itemize}
% \item Case30: A fully-connected neural network with three hidden layers, and the number of neurons for each hidden layer is 32, 16 and 8, respectively.
% \item Case118 and Case200: A fully-connected neural network with three hidden layers, and the number of neurons for each hidden layer is 128, 64 and 32, respectively.
% \end{itemize} 

%\vspace{-0.01in}
\subsection{Performance evaluation}

We show the simulation results in Table~\ref{table2}. 
{For the test cases, we can observe infeasibility if constraint calibrations are not applied during the training stage. With the preventive calibrations, the percentage of feasible solution is improvement, which is up to 38\% (i.e., from 62\% to 100\%). It indicates the effectiveness of the proposed \textsf{DeepOPF+} in ensuring the feasibility of the predicted solutions to the system constraints even with inevitable prediction errors. Also, the differences between the cost of the predicted solutions and that of the reference solutions is minor (at most 1.94\%).} 
Compared with the traditional DC-OPF solver, our \textsf{DeepOPF+} approach reduces the computational time by two orders of magnitude.
Note that the existing DNN-based schemes may not achieve high computational speedups for {both light- and high-loading regime. They may predict infeasible solutions due to violating the operating constraints regarding, e.g., transmission line, and need to resort to post-processing to recover feasible solutions. We show the results of the comparisons for light-load and heavy-load regimes in Appendix~\ref{appendixB}.

Moreover, we observe that a larger calibration magnitude contributes to a higher feasibility rate but larger optimality loss. \rev{It could be interpreted that after constraint calibrations, the DNN approximates the mapping from load inputs to sub-optimal solutions of the adjusted DC-OPF problems.} It indicates the trade-off between ensuring the feasibility and maintaining minor optimality loss of the predicted solutions. We remark the importance of determining the minimal calibration magnitude such that the \textsf{DeepOPF+} scheme can achieve satisfactory speedup performance with minor optimality loss. In our test cases, we found that different cases have different minimal calibration magnitudes. For example, \textsf{DeepOPF+} achieves 100\% feasibility rate for Case30 with a 3.5\% calibration magnitude. For Case118 and Case300, a 5\% calibration magnitude guarantees the 100\% feasibility rates of \textsf{DeepOPF+}. For Case200, with a 7\% calibration magnitude, the predicted solutions of \textsf{DeepOPF+} are all feasible. Under our test case setting, if the constraints calibration magnitude is 7\%, the feasibility percentages of the predicted solutions by \textsf{DeepOPF+} are 100\% over the four test cases. %\rev{ We leave the analysis of characterizing the trade-off between feasibility and optimality and determining the minimal calibration magnitude for future study.}
}

%% file: 7conclusion.tex
\section{Conclusion}\label{sec:conclusion}%\vspace{-0.01in}
In this paper, we propose \textsf{DeepOPF+} as a {preventive} learning approach for solving DC-OPF problems. Our main contributions are that we ensure the feasibility of the predicted solutions by systematically calibrating the constraint limits in the training stage. We theoretically characterize a necessary condition for the magnitude of reduction to guarantee feasibility. Simulation results show that \textsf{DeepOPF+} achieves a computational speed-up by two orders of magnitude as compared to conventional solvers {with minor optimality loss} and ensures feasibility for physical system constraints. Extending the \textsf{DeepOPF+} scheme to solve the general AC-OPF problems is an immediate future direction.

%\vspace{-0.01in}

%% file: 8appendix.tex
%\appendix     
%\vspace{-0.01in}
\begin{appendices}
%%%%%% for the next  appendix section
\section{Proof of Theorem~\ref{theorem1}} \label{apx:proof_of_theorem1}
\vspace{-0.01in}
\begin{proof}
\rev{We prove the two claims one by one. For the first claim, the power flow from $i$-th bus to $j$-th bus is $\frac{1}{x_{ij}}\left( \theta _i-\theta _j \right)$. Given the predicted generation profile, the phase angles can be recovered by the linear relationship (\ref{equation7}).} %allows us to reduce the number of variables to be determined by the DNN. 
Therefore, the power flows on the power network lines are given as\vspace{-0.044in}
\begin{equation*}
PF\hspace{-0.05em}=\hspace{-0.05em} X\left[\begin{array}{c}
0\\
\left( \mathbf{\tilde{B}} \right) ^{-1}\left( \mathbf{\tilde{P}}_{\mathbf{G}}-\mathbf{\tilde{P}}_{\mathbf{D}} \right)
\end{array}
\right]\hspace{-0.05em}=\hspace{-0.05em}\tilde{X}\left( \mathbf{\tilde{B}} \right) ^{-1}\hspace{-0.1em}\left( \mathbf{\tilde{P}}_{\mathbf{G}}-\mathbf{\tilde{P}}_{\mathbf{D}} \right)\hspace{-0.15em},
\end{equation*}
where $X$ is a $|\mathcal{E}|\times N$ matrix and $X_{ai}=x_{ij}, X_{aj}=-x_{ij}$ if there exist a line between $i$-th bus to the $j$-th bus and otherwise $0$. $\tilde{X}$ is the matrix eliminating the column corresponds to the slack bus of matrix $X$. Therefore, the power offsets on each line due to the prediction error of $P_G$ can be expressed as:\vspace{-0.044in}
\begin{equation}
    |\hat{PF}-PF|=\left|\tilde{X}\left( \mathbf{\tilde{B}} \right) ^{-1}(\hat{P}_{\mathbf{G}}-\tilde{P}_{\mathbf{G}})\right|.
\end{equation}%\vspace{-0.03in}
For simplicity, we use $M$ to denote $\tilde{X}\left( \mathbf{\tilde{B}} \right)^{-1}$. {Given the maximal prediction error $\epsilon$, the maximal power offset on the $i$-th line is given as} \vspace{-0.044in}
$$%\max_{i=1, 2, ..., |\mathcal{E}|}
\rev{\sum^{N-1}_{k=1} |M_{ik}|\cdot\epsilon=k_i\cdot \epsilon,}$$
where $k_i=%\max_{i=1, 2, ..., |\mathcal{E}|}
\sum^{N-1}_{k=1} |M_{ik}|$. 

For the second claim, the slack bus generation is given as
\vspace{-0.044in}
\begin{equation}
    P_{G_s}=\sum_{i\in\mathcal{D}}P_{D_i}-\sum_{j\in\mathcal{G},j\neq s}P_{G_j}.
\end{equation}
Given the maximal prediction error $\epsilon$, the maximal power offset on slack bus is given as\vspace{-0.044in}
\begin{equation*}
\max\left|\hat{P_{G_s}}-P_{G_s}\right|=\max\left|\sum_{j\in\mathcal{G},j\neq s}(\hat{P_{G_j}}-P_{G_j})\right|=(|\mathcal{G}-1|)\cdot\epsilon,
\end{equation*}
where $s$ denotes the slack bus index. This completes the proof.
\end{proof}
%\vspace{-0.05in}
% \begin{figure}[!ht]
%     \centering
% 	\includegraphics[width = 0.25\textwidth]{./figs/IEEE30BusSystem.png}
% 	\caption{Topology of the IEEE 30-bus test case.}
% 	\label{fig2}
% \end{figure}

% \section{Topology fof the IEEE 30-bus test case} \label{apx:topology}
% As shown in Fig. \ref{fig2}, the IEEE 30-bus test case consists of 30 buses, 6 generators, 41 branches and 20 loads.

% \begin{figure}[!ht]
% 	\centering
% 	\includegraphics[width = 0.4\textwidth]{./figs/nn_arch.png}
% 	\caption{The detailed architecture of the DNN model for the IEEE 30-bus test case.}
% 	\label{fig3}
% \end{figure}

% \section{The architecture of the DNN model for the IEEE 30-bus test case.} \label{apx:dnnmodel}
% The detailed architecture of the DNN model with input and post-processing stages is shown in Fig. \ref{fig3} for the IEEE 30-bus test case, which includes the dimensions of the input and output as well as the neural network architecture (e.g., the number of neurons on each layer and the corresponding activation function). 

\section{The Speedup comparison for light-load and heavy-load regimes}\label{appendixB}
We also carry out comparative experiments to show the benefits brought by the calibration of constraint limits in the training. More specifically, we compare the speedups performance of the following six schemes:
\begin{itemize}
\item \textsf{DeepOPF}: The previous DNN approach for DC-OPF problem, which uses the default limits in the training and a post-processing procedure to ensure the feasibility of the obtained solutions. 
\item \textsf{DeepOPF+}$\_{V1}$, \textsf{DeepOPF+}$\_{V2}$, \textsf{DeepOPF+}$\_{V3}$, \textsf{DeepOPF+}$\_{V4}$, \textsf{DeepOPF+}$\_{V5}$: The proposed \textsf{DeepOPF+} approach with 0.5\%, 1.5\%, 3.5\%, 5\%, and 7\% line capacity limits and slack bus generation limits calibrations in the training stage, respectively.
% \item \textsf{DeepOPF+}$\_{V2}$: The proposed \textsf{DeepOPF+} approach with 1.5\% line limit and slack bus generation calibration in the training stage.

% \item \textsf{DeepOPF+}$\_{V3}$: The proposed \textsf{DeepOPF+} approach with 3.5\% line limit and slack bus generation calibration in the training stage.
% \item \textsf{DeepOPF+}$\_{V4}$: The proposed \textsf{DeepOPF+} approach with 5\% line limit and slack bus generation calibration in the training stage.
% \item \textsf{DeepOPF+}$\_{V5}$: The proposed \textsf{DeepOPF+} approach with 7\% line limit and slack bus generation calibration in the training stage.
\end{itemize}
Comparative experiments follow the same experimental settings above and the average speedups of the test instances are shown in Table~\ref{table3}. We observe that \textsf{DeepOPF+} improves over existing DNN-based schemes in that it achieves consistent speedups in both light-load and heavy-load regimes, {in which the load data is sampled within $[90\%, 110\%]$ and $[110\%, 130\%]$}, respectively. As mentioned before, existing DNN-based schemes may need a highly computational complexity post-processing procedure to ensure the feasibility in both light- and heavy-load regimes, which is not necessary in the proposed \textsf{DeepOPF+}.

\begin{table}[!th]
	\centering
	\caption{SPEEDUPS PERFORMANCE EVALUATION OF \textsf{DeepOPF+} IN THE LIGHT- AND HEAVY- LOAD REGIMES.}
	\begin{threeparttable}
			\begin{tabular}{c|c|c|c}
			\toprule
			\hline
			\multirow{3}{*}{Case}&
			\multirow{3}{*}{\tabincell{c}{Scheme}} &
			\multicolumn{2}{c}{Average Speedups} \\
			\cline{3-4}
			&&\tabincell{c}{light-load\\regime}&\tabincell{c}{heavy-load\\regime}\\
			\hline
 			\multirow{6}{*}{Case30}&\tabincell{c}{\textsf{DeepOPF}}& $\times$85&$\times$83 \\
			\cline{2-4}
            &\tabincell{c}{\textsf{DeepOPF+}$\_{V1}$}& $\times$85&$\times$83 \\
			\cline{2-4}
			&\tabincell{c}{\textsf{DeepOPF+}$\_{V2}$}& $\times$85&$\times$85 \\
			\cline{2-4}
			&\tabincell{c}{\textsf{DeepOPF+}$\_{V3}$}& $\times$85&$\times$89 \\
			\cline{2-4}
			
			&\tabincell{c}{\textsf{DeepOPF+}$\_{V4}$}& $\times$85&$\times$89 \\
			\cline{2-4}
			&\tabincell{c}{\textsf{DeepOPF+}$\_{V5}$}& $\times$86&$\times$89 \\ %98,51
			\hline
 			\multirow{6}{*}{Case118}&\tabincell{c}{\textsf{DeepOPF}}& $\times$104& $\times$155\\
			\cline{2-4}
            &\tabincell{c}{\textsf{DeepOPF+}$\_{V1}$}& $\times$103 &$\times$191\\
            \cline{2-4}       
			&\tabincell{c}{\textsf{DeepOPF+}$\_{V2}$}& $\times$103&$\times$247 \\%134,75
			\cline{2-4}
			&\tabincell{c}{\textsf{DeepOPF+}$\_{V3}$}& $\times$104&$\times$252 \\
			\cline{2-4}
			&\tabincell{c}{\textsf{DeepOPF+}$\_{V4}$}& $\times$103&$\times$251 \\
			\cline{2-4}
			
			&\tabincell{c}{\textsf{DeepOPF+}$\_{V5}$}& $\times$166&$\times$254 \\
			\cline{2-4}
           
           \hline \multirow{6}{*}{Case200}&\tabincell{c}{\textsf{DeepOPF}}&$\times$57 &$\times$105 \\
			\cline{2-4}
            &\tabincell{c}{\textsf{DeepOPF+}$\_{V1}$}&$\times$60 &$\times$109 \\
            \cline{2-4}        
            
            &\tabincell{c}{\textsf{DeepOPF+}$\_{V2}$}& $\times$67&$\times$138 \\
			\cline{2-4}
			&\tabincell{c}{\textsf{DeepOPF+}$\_{V3}$}& $\times$114&$\times$177 \\
			\cline{2-4}
			&\tabincell{c}{\textsf{DeepOPF+}$\_{V4}$}& $\times$135&$\times$208 \\
			\cline{2-4}
			&\tabincell{c}{\textsf{DeepOPF+}$\_{V5}$}& $\times$141&$\times$208  \\ %233,19
			\hline
			
			\multirow{6}{*}{Case300}&\tabincell{c}{\textsf{DeepOPF}}&$\times$65 &$\times$70 \\
			\cline{2-4}
            &\tabincell{c}{\textsf{DeepOPF+}$\_{V1}$}&$\times$79 &$\times$84 \\
            \cline{2-4}        
            
            &\tabincell{c}{\textsf{DeepOPF+}$\_{V2}$}& $\times$131&$\times$94 \\
			\cline{2-4}
			&\tabincell{c}{\textsf{DeepOPF+}$\_{V3}$}& $\times$143&$\times$112 \\
			\cline{2-4}
			&\tabincell{c}{\textsf{DeepOPF+}$\_{V4}$}& $\times$148&$\times$132 \\
			\cline{2-4}
			&\tabincell{c}{\textsf{DeepOPF+}$\_{V5}$}& $\times$145&$\times$140  \\ %233,19
			\hline
			\bottomrule
		\end{tabular}
% 		\begin{tablenotes}
% 			\footnotesize
% 			\item[*] Speedups are the average of the speedups for each individual test item.
% 		\end{tablenotes}
	\end{threeparttable}
	\label{table3}
\end{table}
\end{appendices}